\documentclass[12pt]{article}
\usepackage{geometry}                % See geometry.pdf to learn the layout options. There are lots.
\usepackage{graphicx}
\usepackage{bm}
\usepackage{amssymb,amsmath,amsthm}
\usepackage{mathrsfs} %script fonts
\def\TODAY{27 January 1989}
\title{\bf Traversable wormholes from surgically modified Schwarzschild spacetimes}
\author{Matt Visser}
\author{{\Large Matt Visser}\\[10pt]
Theoretical Division\\
         T--8, Mail Stop B--285\\
         Los Alamos National Laboratory\\
         Los Alamos, New Mexico 87545\\[10pt]
Present address: \\
School of Mathematics, Statistics, and Computer Science\\
Victoria University of Wellington, New Zealand\\[5pt]
{\sf \small matt.visser@mcs.vuw.ac.nz}  }
\date{\TODAY;  \LaTeX-ed  \today}                                           
\begin{document}
%------------------------------------------------------------------------------------------------------------------------------------------
\maketitle
%------------------------------------------------------------------------------------------------------------------------------------------
% very standard definitions
%------------------------------------------------------------------------------------------------------------------------------------------
\def\d{{\mathrm{d}}}
\newcommand{\scri}{\mathscr{I}}
\newcommand{\sun}{\ensuremath{\odot}}
\def\J{{\mathscr{J}}}
\def\sech{{\mathrm{sech}}}
\newtheorem{theorem}{Theorem}
\def\R{\Re}
\def\throat{\curl\Omega}
\def\ie{\emph{i.e.}}
\def\eg{\emph{e.g.}}
\def\curl{\partial}
\def\infinity{\infty}
%--------------------------------------------------------------------------------------------------------------
\begin{abstract}
%--------------------------------------------------------------------------------------------------------------

In this paper I present a new class of traversable wormholes.  This is done
by surgically grafting two Schwarzschild spacetimes together in such a way
that no event horizon is permitted to form.  This surgery concentrates a
non--zero stress--energy on the boundary layer between the two
asymptotically flat regions.  I shall investigate this stress--energy in
detail using the ``junction condition'' formalism.  A feature of the
present analysis is that this class of traversable wormholes is
sufficiently simple for a (partial) dynamical stability analysis to be
carried out.  The stability analysis places constraints on the equation of
state of the exotic matter that comprises the throat of the wormhole.

\vskip 5pt


\vskip 5 pt


\vskip 5 pt

Nuclear Physics {\bf B328} (1989) 203--212.

\vskip 5 pt

DOI:10.1016/0550-3213(89)90100-4

\enlargethispage{20pt}

%--------------------------------------------------------------------------------------------------------------
\end{abstract}
%--------------------------------------------------------------------------------------------------------------

\clearpage

%%%%%%%%%%%%%%%%%%%%%%%%%%%%%%%%%%%%%%%%%%%%%%%%%%%%%%%%%%%%%%%%%%%%%%%%%%%%%%
% definitions peculiar to this paper
%%%%%%%%%%%%%%%%%%%%%%%%%%%%%%%%%%%%%%%%%%%%%%%%%%%%%%%%%%%%%%%%%%%%%%%%%%%%%%
\def\Ed#1{\par{\bf Editorial Note:} #1 \par}
%%%%%%%%%%%%%%%%%%%%%%%%%%%%%%%%%%%%%%%%%%%%%%%%%%%%%%%%%%%%%%%%%%%%%%%%%%%%%%
\font\Bigfont=cmmi10 scaled\magstep3      %computer modern math italic!
\def\K{\hbox{$\textfont1=\Bigfont \kappa$}} %Big Kappa!
\def\barK{\hbox{$\textfont1=\Bigfont \bar\kappa$}} %Big Kappa!
\def\z{\hbox{$\textfont1=\Bigfont \zeta$}}  %Big Zeta!
%%%%%%%%%%%%%%%%%%%%%%%%%%%%%%%%%%%%%%%%%%%%%%%%%%%%%%%%%%%%%%%%%%%%%%%%%%%%%%
\def\t{\textstyle}
\def\factor{ {\t{1-{2M\over r}}} }
\def\Factor{ {\t{1-{2M\over a}}} }
\def\FFactor{ {\t {1-{2M\over a}+{\dot a}^2} } }
\def\RNfactor{ {\t {1-{2M\over r}+{Q^2\over r^2}} } }
\def\RNFactor{ {\t {1-{2M\over a}+{Q^2\over a^2}} } }
\def\dsqOmega{ d\theta^2 + \sin^2\theta\;d\phi^2 }
\def\M{{$\script M$}}
\def\half{{1\over2}}
%%%%%%%%%%%%%%%%%%%%%%%%%%%%%%%%%%%%%%%%%%%%%%%
 
%%%%%%%%%%%%%%%%%%%%%%%%%%%%%%%%%%%%%%%%%%%%%%%%%%%%%%%%%%%%%%%%%%%%%%%%%%%%%%
\section{Introduction.}
\setcounter{equation}{0}%
%%%%%%%%%%%%%%%%%%%%%%%%%%%%%%%%%%%%%%%%%%%%%%%%%%%%%%%%%%%%%%%%%%%%%%%%%%%%%%
 
Recently there has occurred a major Renaissance in wormhole physics.  While
most energies are being foucused on wormholes as possibly significant
contributions to quantum gravity, I feel that the analysis of classical
traversable wormholes merits serious attention.  Analyses of traversable
wormholes have recently been presented by Morris and Thorne~\cite{MT}, by Morris, Thorne, and Yurtsever~\cite{MTY}, and by the present author~\cite{surgical}.  Traversable wormholes are
specifically {\sl designed} so that they may in principle be used by humans
to travel between universes (or distant parts of the same universe) without
fatal effects on the traveller.  A major result of these investigations is
that violations of the weak energy hypothesis are guaranteed to occur at
the throat of a traversable wormhole.
 
In this paper I present a new class of traversable wormholes.  In an
earlier publication, I discussed a class of spherically asymmetric
traversable wormholes~\cite{surgical}.  In this paper I shall pursue a
different topic.  I shall adopt the constraint of spherical symmetry, and
shall assume that all matter is confined to a thin boundary layer between
universes.  Thus the models I consider are a subset of the ``absurdly
benign'' wormholes of reference~\cite{MT}.  The models are constructed by by
surgically grafting two Schwarzschild spacetimes together in such a way
that no event horizon is permitted to form.  This surgery (naturally)
concentrates a non--zero stress--energy on the boundary layer between the
two universes.  We shall investigate this stress--energy in detail using
the ``junction condition'' formalism (a.k.a. the ``boundary layer''
formalism).  A major innovation in the present analysis is that these
wormholes are sufficiently simple for a (partial) dynamical stability
analysis to be carried out.  Such a stability analysis was totally
unmanageable in the models considered in references~\cite{MT, MTY, surgical}.  The
generalization to traversable wormholes based on surgically modified
Reissner--Nordstr\"om spacetimes is immediate.
 
It is most illuminating to first construct the class of models to be
considered in the static case, ignoring stability questions.  Once details
of the static case have been spelled out, I shall turn attention to the
dynamical analysis of stability.  I shall be limited to considering
spherically symmetric motions of the wormhole throat.  The technical
details of this analysis will closely parallel that of Blau, Guendelman,
and Guth~\cite{BGG}, though they were looking at a totally
different physical system.
 
The static analysis already enforces the presence of ``exotic
stress--energy'' (\ie, violations of the weak energy hypothesis~\cite{MT, MTY, surgical}).  The stability analysis places constraints on the
equation of state of this exotic stress--energy.  Suitable candidates for
the equation of state of this exotic stress--energy are identified.  If the
wormhole is to be absolutely stable, (rather than metastable), then the
gravitational mass as measured at spatial infinity must be non--positive.
This behavior is alarming, but not necessarily fatal.
 
%%%%%%%%%%%%%%%%%%%%%%%%%%%%%%%%%%%%%%%%%%%%%%%%%%%%%%%%%%%%%%%%%%%%%%%%%%%%%%
\section{Schwarzschild surgery.}
\setcounter{equation}{0}%
%%%%%%%%%%%%%%%%%%%%%%%%%%%%%%%%%%%%%%%%%%%%%%%%%%%%%%%%%%%%%%%%%%%%%%%%%%%%%%
 
To construct the wormholes of interest, consider the ordinary Schwarzschild
solution to the vacuum Einstein field equations:
\begin{equation}
ds^2 = -(\factor)dt^2 + {dr^2\over(\factor)} +r^2(\dsqOmega).
\end{equation}
I utilize the ordinary Scwarzschild coordinates, and do {\sl not}
maximally extend the manifold, as that would prove to be unprofitable.  Now
take {\sl two} copies of this manifold, and remove from them the
four--dimensional regions described by $\Omega_{1,2} \equiv \{ r_{_{1,2}}
\leq a \;|\; a > 2M \} $.  One is left with two geodesically incomplete
manifolds with boundaries given by the timelike hypersurfaces
$\curl\Omega_{1,2} \equiv \{ r_{_{1,2}} = a \;|\; a > 2M \}$.  Now identify
these two timelike hypersurfaces (\ie,\ $\curl\Omega_1 \equiv
\curl\Omega_2$).  The resulting spacetime {$\mathscr M$} is geodesically
complete and possesses two asymptotically flat regions connected by a
wormhole.  The throat of the wormhole is at $\curl\Omega$.  Because
{$\mathscr M$} is piecewise Schwarzschild, the stress--energy tensor is
everywhere zero, except at the throat itself.  At $\curl\Omega$ one expects
the stress--energy tensor to be proportional to a delta function.  This
situation is made to order for an application of the ``junction condition''
formalism, also known as the ``boundary layer'' formalism~\cite{BGG, MTW}.  The nature and behavior of the stress--energy tensor is
the major focus of this paper.  Note that the class of traversable
wormholes I have just described is a subclass of the ``absurdly benign''
wormholes of reference 1, obtained in the limit where the thickness of the
transition region containing exotic stress--energy shrinks to zero.
Finally, note that the condition $a > 2M$ is necessary to prevent the
formation of an event horizon.
 
Since all the stress--energy is concentrated on the throat, the throat may
be viewed as behaving like a domain wall between the two universes.  The
simplest domain wall one can construct is the classical membrane, but this
will be shown to be unstable.  More generally one may consider a domain
wall consisting of a membrane that has some (2+1)--dimensional matter
trapped on it.  Domain walls of this type can in principle possess
essentially arbitrary equations of state.  We shall use the stability
analysis to constrain the equation of state.
 
I also wish to mention that the analysis soon to be presented generalizes
immediately to traversable wormholes based on surgical modifications of the
Reissner--Nordstr\"om spacetime.  Merely repeat the above discussion using
the metric
\begin{equation}
ds^2 = -\left(\RNfactor\right)dt^2 + {dr^2\over\left(\RNfactor\right)}
        +r^2\left(\dsqOmega\right).
\end{equation}

%%%%%%%%%%%%%%%%%%%%%%%%%%%%%%%%%%%%%%%%%%%%%%%%%%%%%%%%%%%%%%%%%%%%%%%%%%%%%%
\section{Junction conditions and the Einstein equations.}
\setcounter{equation}{0}%
%%%%%%%%%%%%%%%%%%%%%%%%%%%%%%%%%%%%%%%%%%%%%%%%%%%%%%%%%%%%%%%%%%%%%%%%%%%%%%
 
I now briefly review the junction formalism~\cite{BGG, MTW}.  Consider two
four--dimensional spacetimes $\Omega_{1,2}$ with boundaries
$\curl\Omega_{1,2}$.  We wish to join these manifolds at their boundaries
to create a new manifold $\Omega = \Omega_1 \oplus \Omega_2$ that has no
boundary.  The first junction condition is that the three--dimensional
geometries of $\curl\Omega_1$ and $\curl\Omega_2$ must be the same.  (That
is, the first fundamental forms of $\curl\Omega_1$ and $\curl\Omega_2$ must
be identical.)  This means that we wish the geometry of $\Omega$ to at
least be continuous at the junction.  However, the metric need not be
differentiable at the junction, so the affine connection may be
discontinuous there, and the Riemann tensor may possess a delta function
singularity there.  This statement may be quantified in terms of the second
fundamental forms of the boundaries $\curl\Omega_{1,2}$.  Let us adopt
Riemann normal coordinates at the junction.  Let $\eta$ denote a coordinate
normal to the junction, with $\eta$ positive in $\Omega_1$ and negative in
$\Omega_2$.  The second fundamental forms are then
\begin{equation}
K^i{}_j{}^\pm = {\t\half} \; g^{ik} \cdot \left.{\curl
g_{kj}\over\curl\eta}\right|_{\eta=\pm0.}
\end{equation}
The Ricci tensor at the junction is easily calculated in terms of the
discontinuity in the second fundamental forms.  Define $\K_{ij} =
K_{ij}{}^+ - K_{ij}{}^-$, then
\begin{equation}
R^\mu{}_\nu(x) = \left[ \begin{matrix} \K^i{}_j(x) & 0\\ 0&\K(x)\end{matrix} \right]
\cdot \delta(\eta) + R^\mu{}_\nu{}^+(x) \cdot \Theta(\eta)
 + R^\mu{}_\nu{}^-(x) \cdot \Theta(-\eta)
\end{equation}
(We adopt the convention that $\Theta(0)=0$.) The stress--energy tensor is
\begin{equation}
T^{\mu\nu}(x) = S^{\mu\nu}(x) \cdot \delta(\eta)
+T^{\mu\nu+}(x)\cdot\Theta(\eta)
+T^{\mu\nu-}\cdot\Theta(-\eta).
\end{equation}
Conservation of stress--energy severely constrains
the surface stress--energy.  It is easy to see that $S^{\eta\eta}
\equiv 0\equiv S^{i\eta}$, so that the only nonzero components of $S$ are
the $S^i{}_j$.  Stress--energy may be exchanged between the surface layer
at the junction and its surroundings, subject to the constraint
$S^{ij}{}_{|j} = - T^{i\eta+} + T^{i\eta-}$.  Finally the condition of
pressure balance reads $\barK_{ij} S^{ij} = T^{\eta\eta+} -
T^{\eta\eta-}$, where $\barK_{ij} = \half\{ K_{ij}{}^{+} +
K_{ij}{}^{-}\}$.  The Einstein field equations lead to an expression for
the surface stress--energy tensor
\begin{equation}
S^i{}_j = -{\t{c^4\over8\pi G}} \cdot \left[\K^i{}_j - \delta^i{}_j \;
\K^k{}_k \right]_.
\end{equation}
 
This completes the review of the general case.  For the spherically
symmetric and reflection symmetric case at hand considerable
simplifications occur.  Firstly, $K^{+}=-K^{-}=\half\K$, while
$\barK\equiv0$.  Secondly $T^\pm \equiv0$, so the pressure balance
constraint is vacuous, while for the surface stress--energy $S^{ij}{}_{|j}
= 0$.  Thirdly, spherical symmetry implies that
\begin{equation}
\K^i{}_j = \left[ \begin{matrix}\K^\tau{}_\tau&0&0\\
                            0&\K^\theta{}_\theta&0\\
                            0&0&\K^\theta{}_\theta\end{matrix} \right]_,
\end{equation}
while the surface stress--energy tensor may be written in terms of the
surface energy density $\sigma$ and surface tension $\vartheta$ as
\begin{equation}
S^i{}_j = \left[ \begin{matrix}-\sigma&0&0\\
                            0&-\vartheta&0\\
                            0&0&-\vartheta\end{matrix} \right]_.
\end{equation}
Adopting ``Geometrodynamic units'' ($G\equiv 1\equiv c$), the field
equations become
\begin{equation}
\sigma = -{\t{1\over4\pi}}\cdot \K^\theta{}_\theta; \qquad
\vartheta = -{\t{1\over8\pi}}\cdot \{\K^\tau{}_\tau +
\K^\theta{}_\theta \}.
\end{equation}
This has now reduced the computation of the stress--energy tensor to that of
computing the two non--trivial components of the second fundamental form.
This is very easy to do in the static case, and still quite manageable if
the throat is in motion.

%%%%%%%%%%%%%%%%%%%%%%%%%%%%%%%%%%%%%%%%%%%%%%%%%%%%%%%%%%%%%%%%%%%%%%%%%%%%%%
\section{Static wormholes.}
\setcounter{equation}{0}%
%%%%%%%%%%%%%%%%%%%%%%%%%%%%%%%%%%%%%%%%%%%%%%%%%%%%%%%%%%%%%%%%%%%%%%%%%%%%%%
 
The static case is particularly simple. We note
\begin{equation}
K^i{}_j{}^\pm = g^{ik} \cdot \left.{\curl g_{kj}\over\curl\eta}
                    \right|_{r=a}
                   = \left.{\curl r\over \curl\eta}\right|_{r=a} \cdot
                      g^{ik} \cdot
                    \left.{\curl g_{kj}\over\curl r }\right|_{r=a.}
\end{equation}
Now ${\curl r\over \curl\eta} = \sqrt{\factor}$, so that
\begin{equation}
K^\tau{}_\tau{}^\pm = \pm{{M\over a^2}\over\sqrt{\Factor}}; \qquad
K^\theta{}_\theta{}^\pm = \pm{\sqrt{\Factor}\over a}.
\end{equation}
Which immediately leads to
\begin{equation}
\sigma = -{\t{1\over2\pi a}} \cdot \sqrt{\Factor}; \qquad \qquad
\vartheta = -{\t{1\over4\pi a}} \cdot { {1-{M\over a}} \over
              \sqrt{\Factor}}.
\end{equation}
 
Note that the energy density is negative.  This is just a special case of
the de--focusing arguments presented in references~\cite{MT, MTY, surgical}
\ whereby the presence at the throat of ``exotic'' matter (matter that
violates both the weak energy hypothesis and the averaged weak energy
hypothesis) was inferred.  While somewhat alarming, should not cause too
much consternation.  It cannot be strongly enough emphasised that the weak
energy hypothesis has been experimentally tested in the laboratory, and has
been experimentally shown to be {\sl false}.  The averaged weak energy
hypothesis is more subtle to test experimentally, and no conclusive results
can presently be drawn.  It is not commonly appreciated, but it is in fact
true, that the observation of the Casimir effect between parallel
plates~\cite{Casimir} is sufficient to experimentally disprove the
weak energy hypothesis, (and also disprove the strong and dominant energy
hypotheses).  For analyses of the form of the stress--energy tensor between
parallel plates see Gibbons~\cite{Gibbons} and
DeWitt~\cite{DeWitt}.  Further comments along these lines
may be found in Roman~\cite{Roman}, and in the seminal
articles by Morris and Thorne~\cite{MT}, and Morris, Thorne, and Yurtsever~\cite{MTY}.  In this regard it is perhaps somewhat embarrassing to realise that
the experimental observations disproving the (unaveraged) energy hypotheses
predate the formulation of the (unaveraged) energy hypotheses by some 25
years.  It is an open question as to whether or not quantum theory
satisfies the averaged weak energy hypothesis, so until this question is
settled the existence of this class of traversable wormholes should be
viewed with some caution.  In the meantime, it would appear that the best
prospects for a theoretical understanding of exotic stress--energy are
within the context of semiclassical quantum gravity.  The surface tension
is also negative, but this merely implies that we are dealing with a
surface {\sl pressure}, not a tension.  It should not be too surprising
that a positive pressure is needed to prevent collapse of the wormhole
throat.
 
Two special cases are of immediate interest: \\
 {\sl ------ The
classical membrane}, described by the three--dimensional generalization of
the Nambu--Goto action, satisfies the equation of state $\sigma =
\vartheta$.  In this case $a=3M$ and 
\begin{equation}
\sigma = \vartheta = - {1\over2\pi a} \cdot {1\over\sqrt3}.
\end{equation}
This should be compared with the analysis in reference \cite{surgical}, wherein
negative tension classical strings were used to construct spherically
asymmetric wormholes with polyhedral throats.  When we turn to the
dynamical analysis, we shall quickly see that this type of wormhole is
dynamically unstable. 
\\ {\sl ------ Traceless stress--energy}.  The
case $S^k{}_k = 0$ (\ie,\ $\sigma+2\vartheta=0$) is of interest because it
describes massless stress--energy confined to the throat.  (Such a
stress--energy tensor arises from considering the Casimir effect for
massless fields, a popular way of obtaining exotic stress--energy~\cite{MT, MTY, surgical}.)  Unfortunately, in this case no solution to the
Einstein field
equations exists.  ($\sigma,\vartheta$ prove to be imaginary.)  This result
is rather depressing as it indicates that consideration of the Casimir
effect associated with {\sl massless} fields is rather less useful than
expected.
 
Following the earlier discussion, the analysis immediately generalizes.
Observe that for a traversable wormhole based on surgical modification of
the Reissner--Nordstr\"om spacetime:
\begin{equation}
\sigma = - {\t{1\over2\pi a}}  \cdot \sqrt\RNFactor; \qquad
\vartheta = - {\t{1\over4\pi a}} \cdot{ {1-{M\over a}} \over \sqrt\RNFactor }.
\end{equation}

%%%%%%%%%%%%%%%%%%%%%%%%%%%%%%%%%%%%%%%%%%%%%%%%%%%%%%%%%%%%%%%%%%%%%%%%%%%%%%
\section{Dynamic wormholes.}
\setcounter{equation}{0}%
%%%%%%%%%%%%%%%%%%%%%%%%%%%%%%%%%%%%%%%%%%%%%%%%%%%%%%%%%%%%%%%%%%%%%%%%%%%%%%
 
To analyse the dynamics of the wormhole, we permit the radius of the throat
to become a function of time $a \mapsto a(t)$.  Note that by an application
of Birkhoff's theorem we can be confident that at large radius [in fact for
any $r > a(t) $] the geometry will remain that of a piece of Schwarzschild
spacetime (or Reissner--Nordstr\"om spacetime).  In particular, the assumed
spherical symmetry is a sufficient condition for us to conclude that there
is no gravitational radiation regardless of the behavior of $a(t)$.  Let
the position of the throat of the wormhole be described by $x^\mu (t,
\theta, \phi) \equiv (t, a(t), \theta, \phi)$, so that the four velocity of
a piece of stress--energy at the throat is:
\begin{equation}
U^\mu = \left({\t {dt\over d\tau}, {da\over d\tau}, 0, 0 } \right)
      = \left({ \sqrt{1-{2M\over a} + {\dot a}^2 }\over\Factor },
        \;\; \dot a, \;\; 0, \;\; 0 \right).
\end{equation}
The unit normal to $\curl\Omega$ is computed to be
\begin{equation}
\xi^\mu = \left( {\dot a \over \Factor}, \;\; \sqrt{\Factor +
{\dot a}^2},
                 \;\; 0, \;\; 0
          \right).
\end{equation}
 
The $\theta\theta$ and $\phi\phi$ components of the second fundamental form
are
\begin{equation}
K^\theta{}_\theta \equiv K^\phi{}_\phi =
   \left. {1\over r} \cdot {\curl r\over \curl\eta} \right|_{r=a} =
   {1\over a} \cdot \sqrt{\FFactor}.
\end{equation}
Evaluating the $\tau\tau$ component of $K$ is more difficult.  One may,
naturally, proceed via brute force.  It is more instructive to present a
short digression.  We note that
\begin{eqnarray}
K^\tau{}_\tau = -K_{\tau\tau} &=& \xi_{\tau;\tau} \nonumber\\
                                &=& - U^\mu U^\nu \xi_{\nu;\mu} \nonumber\\
                                &=& + U^\nu \xi_\mu U^\mu{}_{;\nu} \nonumber\\
                                &=& \xi_\mu (U^\mu{}_{;\nu} U^\nu) \nonumber\\
                                &=& \xi_\mu A^\nu.
\end{eqnarray}
Here $A^\mu$ is the four--acceleration of the throat.  By the spherical
symmetry of the problem, the four--acceleration is proportional to the unit
normal $A^\mu \equiv A \cdot \xi^\mu$, so that $K^\tau{}_\tau = A \equiv$
{\sl magnitude of the four--acceleration}.  To evaluate the
four--acceleration,
utilize the fact that $k^\mu \equiv ({\curl\over\curl t})^\mu \equiv
(1,0,0,0)$ is a Killing vector for the underlying Schwarzschild geometry.
Note that $k_\mu = \left( -(\Factor),0,0,0\right)$, so that $\dot a =
-k_\mu \xi^\mu$, and $\sqrt{\FFactor} = -k_\mu U^\mu$.  With considerable
inspired guess--work and hindsight, it becomes interesting to evaluate:
\begin{eqnarray}
{\t{D\over D\tau}}(k_\mu U^\mu) &=& k_{\mu;\nu} \; U^\nu \; U^\mu +
                              k_\mu \; {\t{DU^\mu\over D\tau}}  \nonumber\\
&=& k_\mu \; A \; \xi^\mu  \nonumber\\
&=& -A \; \dot a. 
\end{eqnarray}
On the other hand
\begin{eqnarray}
{\t{D\over D\tau}}(k_\mu U^\mu)
&=& -{\t{D\over D\tau}}\sqrt{\FFactor}  \nonumber\\
&=& -{1\over\sqrt{\FFactor}} \cdot \left({\t{M\over a^2}} + \ddot a \right)
\cdot\dot a. 
\end{eqnarray}
Comparing the two calculations, we find that the four--acceleration of
the throat is
\begin{equation}
A = {\left( \ddot a + {M\over a^2} \right) \over \sqrt\FFactor }.
\end{equation}
 
The Einstein field equations become
\begin{equation}
\label{E:Einstein}
\sigma = -{\t{1\over2\pi a}} \cdot \sqrt\FFactor; \qquad
\vartheta = -{\t{1\over4\pi a}} \cdot
          { \left( 1 -{M\over a} + {\dot a}^2 + a \ddot a \right) \over
             \sqrt\FFactor}.
\end{equation}
It is relatively easy to check that equations \ref{E:Einstein} imply the
conservation of stress-energy
\begin{equation}
\dot \sigma = -2 (\sigma-\vartheta) {\dot a\over a}, \qquad \hbox{that is:}
\qquad
{\t{D\over D\tau}} (\sigma a^2) = \vartheta\cdot {\t{D\over D\tau}} (a^2).
\end{equation}
As is usual, there is a redundancy between the Einstein field equations and
the covariant conservation of stress--energy. With the field equations for
a moving throat in hand, the dynamical stability analysis will prove
simple.
 
%%%%%%%%%%%%%%%%%%%%%%%%%%%%%%%%%%%%%%%%%%%%%%%%%%%%%%%%%%%%%%%%%%%%%%%%%%%%%%
\section{Stability Analysis.}
\setcounter{equation}{0}%
%%%%%%%%%%%%%%%%%%%%%%%%%%%%%%%%%%%%%%%%%%%%%%%%%%%%%%%%%%%%%%%%%%%%%%%%%%%%%%
 
The Einstein equations obtained in the previous section may be recast as
the pair
\begin{equation}
{\dot a}^2 -{\t{2M\over a}} - (2\pi\sigma a)^2 = -1; \qquad
\dot \sigma = -2 (\sigma-\vartheta) {\dot a\over a}.
\label{E:energy}
\end{equation}
 
Consider the classical membrane.  The equation of state is $\sigma =
\vartheta$, so that $\dot\sigma \equiv 0$.  It is immediately clear from
equation \ref{E:energy} that a traversable wormhole built using a classical
membrane is dynamically unstable.  We need merely observe that the
potential in \ref{E:energy} is unbounded below.  Wormholes of this type either
collapse to $a=0$ or blow up to $a=\infty$ depending on the initial
conditions.  For example, if $a\gg M$ we may write down the approximate
solution $a(\tau) = {1\over2\pi\sigma}\cosh(2\pi\sigma\tau)$.  Even the
somewhat {\sl outr\'e} condition that $M<0$ will only help to stabilize the
wormhole against collapse, it will do nothing to prevent the system
exploding.  It should be noted that since the surface energy density is
already negative, the possibility of a negative total gravitational mass is
no longer excluded.  Since the presence of the wormhole has allowed us to
excise the otherwise naked singularity at $r=0$, this geometry does not
violate the cosmic censorship hypothesis even for $M<0$.  Furthermore, even
if $M<0$, one still must require $2\pi \cdot a^{3/2} \cdot \sigma(a) \to
k_0 < \sqrt{2 |M|}$ as $a\to0$ in order for the surface density term to not
overwhelm the mass term.
 
More generally note that for $M>0$ the potential near $a=0$ is unbounded
below, regardless of the behavior of $\sigma(a)$. Indeed if $a<2M$, we see
that a runaway solution develops with $a\to0$ as $\tau\to\infinity$.
Physically this is a reflection of the fact that if the throat falls within
its own Schwarzschild radius, then the wormhole is doomed. Thus if $M>0$
the best we can hope for is that the wormhole be {\sl metastable} against
collapse to $a=0$.
 
If we consider the behavior as $a\to\infinity$, we see that the wormhole
is stable against explosion if and only if $2\pi \cdot a \cdot \sigma(a)
\to k_\infinity < 1$.  If this condition is violated, the wormhole is at
best {\sl metastable}.  This condition on $\sigma(a)$ will be shown to
imply a constraint on the equation of state of the domain wall in the
region $\sigma \approx 0$.

%%%%%%%%%%%%%%%%%%%%%%%%%%%%%%%%%%%%%%%%%%%%%%%%%%%%%%%%%%%%%%%%%%%%%%%%%%%%%%
\section{Equation of state.}
\setcounter{equation}{0}%
%%%%%%%%%%%%%%%%%%%%%%%%%%%%%%%%%%%%%%%%%%%%%%%%%%%%%%%%%%%%%%%%%%%%%%%%%%%%%%
 
To constrain the equation of state, we use the fact that $2\pi \cdot a
\cdot \sigma(a) \to k_\infinity <1$ as $a \to \infinity$.  Since this
implies that $\sigma(a) \to 0$ at spatial infinity, it becomes interesting
to expand the equation of state in a Taylor series around $\sigma = 0$:
\begin{equation}
\vartheta(\sigma) = \vartheta_0 + k^2 \sigma + o(\sigma^2).
\end{equation}
 
From this assumed equation of state, and the conservation of
stress--energy, one may calculate $\sigma(a)$.  Specifically, ignoring
$o(\sigma^2)$ terms
\begin{equation}
\vartheta_0 + k^2 \sigma = \vartheta = \sigma + {\t{1\over2}} a
                        {\t{d\sigma\over da}}.
\end{equation}
This differential equation is easily solved
\begin{equation}
\sigma(a) = {\vartheta_0\over 1-k^2} + \sigma_0 \cdot
(a/a_0)^{2(k^2 -1)}.
\end{equation}
By looking at the $a\to \infinity$ behavior we see that $\vartheta_0 = 0$
and $k^2 \leq {1\over2}$. On the other hand, since equations \ref{E:Einstein} 
imply that $\sigma$ and $\vartheta$ have the same sign, $k^2 \geq 0$. Thus
\begin{equation}
\vartheta(\sigma) = k^2 \sigma + o(\sigma^2); \qquad
k \in [0, {\t{1\over\sqrt2}}].
\end{equation}
In particular, the case $k=0$, (corresponding to negative energy dust), is
stable against explosion. Dust is, however, unstable against collapse. 
To stabilize the wormhole requires either (i) $M=0$ and $k \geq \sqrt{1/2}$, or (ii) $M<0$ and $k> 1/2$, or (iii) $M<0$, $k=1/2$, and $2\pi \sigma_0 a_0^{3/2} < \sqrt{2|M|}$. So for the case $M=0$ we are uniquely led to $k=1/\sqrt{2}$, $\vartheta={1\over2} \sigma$, $\sigma= \sigma_0 a_0/a$; while for $M<0$ the entire range $k \in (1/2, 1/\sqrt{2}]$ is acceptable, corresponding to the range of ``semi-soft'' equations of state between $\vartheta={1\over4}\sigma$ and $\vartheta={1\over2}\sigma$.
%%%%%Rewrite.
%If $M<0$ then stability against collapse requires that $k \in [1/2,
%\infinity)$. In this case the region $k \in [1/2, 1/\sqrt{2}]$ gives rise to
%an equation of state describing a system that is stable against both
%collapse and explosion. 
A slightly more general
form of the equation of state, stable against both
collapse and explosion, is %the two phase system:
\begin{equation}
\vartheta(\sigma) = k^2_+ \cdot \sigma\cdot \Theta(\sigma_0-\sigma) +
                    k^2_- \cdot \sigma\cdot \Theta(\sigma-\sigma_0).
\end{equation}
(Note the discontinuity at $\sigma = \sigma_0$.) This equation of state
leads to a surface density
\begin{equation}
\sigma(a) = \sigma_0 \left\{ \left({{ a\over a_0}}\right)^{2(k_+^2-1)}
\cdot \Theta(a-a_0) +
                             \left({{ a\over a_0}}\right)^{2(k_-^2-1)}
\cdot \Theta(a_0-a) \right\}_.
\end{equation}
Stability against explosion requires $k_+ \in [0,{1\over\sqrt2}]$, while stability against collapse requires  $k_- \in [{1\over2},\infty)$. (If $k_-={1\over2}$, then one also requires $2\pi\sigma_0a_0^{3/2}< \sqrt{2|M|}$.)
%If $M>0$
%the system is at best metastable against collapse. If $M=0$ stability
%against collapse requires $k_- \in [1/\sqrt{2},\infinity)$, while if $M<0$
%we merely require $k_- \in [1/2,\infinity)$.
For example, the equation of state
$\vartheta(\sigma) = \sigma \cdot \Theta(\sigma - \sigma_0)$ leads to a
wormhole that is stable against both explosion and collapse.  This example
behaves like dust for $\sigma < \sigma_0$, and behaves like a
classical membrane for $\sigma > \sigma_0$.
 
The conclusions drawn regarding the equation of state depend critically
only upon the assumed Taylor series expansion for $\vartheta(\sigma)$
around $\sigma=0$.  It is of course possible (though unlikely) that the
equation of state be non--analytic at that point, so that no such expansion
exists.  I have investigated only one example of this type of behavior.
Assume the equation of state is
\begin{equation}
\vartheta(\sigma) = \sigma_0 \cdot (\sigma/\sigma_0)^{1+(\nu/2)},
\end{equation}
Then conservation of stress energy implies that
\begin{equation}
\sigma(a) = \sigma_0 \cdot \left[ 1 + (a/a_0)^\nu  \right]^{-2/\nu}.
\end{equation}
If $\nu$ is positive, then $\sigma \to \sigma_0 (a/a_0)^{-2}$ as $ a\to
\infinity$, so the wormhole is stable against explosion.  Also if $\nu >
0$, then $\sigma \to \sigma_0$ as $a \to 0$, so the wormhole is stable
against collapse for $M \leq 0$.
 
%%%%%%%%%%%%%%%%%%%%%%%%%%%%%%%%%%%%%%%%%%%%%%%%%%%%%%%%%%%%%%%%%%%%%%%%%%%%%%
\section{Conclusions.}
\setcounter{equation}{0}%
%%%%%%%%%%%%%%%%%%%%%%%%%%%%%%%%%%%%%%%%%%%%%%%%%%%%%%%%%%%%%%%%%%%%%%%%%%%%%%
 
In this paper I have described and investigated a new class of traversable
wormholes.  Because of the simple nature of the geometry a partial
stability analysis is possible.  As is usual for traversable wormholes,
exotic stress--energy (violating the weak energy hypothesis) is present at
the throat of the wormhole.  Additionally, (global) stability of the
wormhole against collapse requires that the gravitational mass of the
wormhole as seen by an observer at spatial infinity be non--positive.
Constraints are placed on the behavior of the surface energy density as a
function of the radius of the throat.  These constraints may be transformed
into (limited) statements about the equation of state.
 
The presence of negative energy density and gravitational mass is certainly
disturbing but, given our present lack of knowledge, is not immediately
fatal.  It is encouraging that certain equations of state for the domain
wall lead to stable wormholes.  It would certainly be of interest to study
the hypothesized material comprising the throat in more detail.  If nothing
else we may yet be able to rule out the existence of traversable wormholes
on physical grounds.

%------------------------------------------------------------------------------------------------------------------------------------------

%------------------------------------------------------------------------------------------------------------------------------------------
\end{document}